\documentclass[10pt,twocolumn,floatfix]{revtex4-1}
\usepackage{graphicx}
\usepackage{epstopdf}
\usepackage{amssymb}
\usepackage{mathrsfs}
\usepackage{amsmath}
\usepackage{color}
\usepackage{latexsym}
\usepackage{amsfonts}
\usepackage{hyperref}
\usepackage{subfigure}
\usepackage{wasysym}
\usepackage{dcolumn}
\usepackage{bm}

\usepackage{natbib}

\begin{document}

\selectfont

\title{Contribution of Water to Pressure and Cold Denaturation of Proteins}

\author{Valentino Bianco}

\author{Giancarlo~Franzese}
\affiliation{Departament de F\'isica Fonamental, Universitat de Barcelona, Mart\'i i Franqu\`es 1, 08028 Barcelona, Spain}
\email{gfranzese@ub.edu}

\date{Received 11 May 2015, revised manuscript received 14 July 2015, published 1 September 2015}

\definecolor{corr14okt}{rgb}{0,0,1} 
\definecolor{moved}{rgb}{0,0,0}

\begin{abstract} The mechanisms of cold- and pressure-denaturation of
  proteins are matter of debate and are commonly understood as due
  to water-mediated interactions. 
  Here we study several cases of proteins, with or
without a unique native state, with or without hydrophilic residues,
by means of  a coarse-grain protein model in 
explicit solvent. We show, using  Monte Carlo simulations, 
that taking into account how water at the protein
interface changes its hydrogen bond  properties and its
density fluctuations is enough to predict protein stability regions with elliptic shapes in the
temperature-pressure plane, consistent with previous
theories.
Our results clearly identify the different mechanisms with which water
participates to denaturation and open the perspective to develop
advanced computational design tools for protein engineering.
\\\\\\PACS number
87.15.Cc,
87.15.A-, 	
87.15.kr.
\end{abstract}

\maketitle

Water plays an essential role in driving the folding of a protein
and in stabilizing the tertiary protein
structure in its native state \cite{Levy2006, Kinoshita2009b}.  
Proteins can denaturate---unfolding their structure and loosing their
activity---as a consequence of changes in the environmental conditions. 
Experimental data show that for many proteins
the native folded state
is stable in a limited range of temperatures $T$ and
pressures $P$
  \cite{Ravindra2003, Pastore2007,
   Meersman2000, HummerGerhard1998, MeersmanFilip2006, 
 NucciNathaniel2014} 
and that partial folding is $T$-modulated also in ``intrinsically disordered proteins''
\cite{WuttkeRene2014}. By hypothesizing that proteins have only two different states, folded
($f$) and unfolded ($u$),  and that the $f\longleftrightarrow
u$ process is reversible
at any moment, Hawley proposed a theory 
\cite{hawley1971} that predicts 
a close stability region (SR) with an elliptic shape in the 
$T-P$ plane, consistent with the experimental data \cite{Smeller2002}.

Cold- and $P$-denaturation of proteins have been related to the equilibrium properties of
the hydration water \cite{delosrios2000,
 Tsai2002, Paschek2005,Marques2003,
  Patel2007, Athawale2007, Dias2008, Nettels2009, Best2010, Jamadagni2010, 
Badasyan2011, Matysiak2012}.
However, the interpretations of the
mechanism is still 
controversial \cite{paschekPRL2004_new, paschekPNAS2005_new, sumiPCCP2011,   Coluzza2011, diasPRL2012_new,
  Das2012, Sarma2012, Fernandez2013, FranzeseBiancoFoodBio2013,
  NucciNathaniel2014, Abeln2014, Ben-Naim2014, YangNature2014, RochePNAS2012, sirovetzJPCB2015_new}.
High-$T$
denaturation is easily understood in terms of thermal fluctuations
that disrupt the compact protein conformation: the open protein
structure increases the entropy $S$ minimizing the global Gibbs free
energy $G\equiv H-TS$, where $H$ is the total enthalpy.
High-$P$ unfolding can be explained by the loss of
  internal cavities in the folded states of proteins
  \cite{RochePNAS2012},
  while denaturation at negative $P$ has been experimentally
  observed \cite{Larios2010} and simulated \cite{Larios2010,
    HatchJPCB2014} recently. 
Cold- and
$P$-unfolding can be thermodynamically justified assuming an
enthalpic gain of the solvent upon denaturation process, without
specifying the origin of this gain from molecular interactions
\cite{muller1990}.  
Here, we propose a molecular-interactions model for proteins solvated
by explicit water, based on the 
 ``many-body'' water model \cite{FMS2003, Stokely2010,
  MSPBSF2011,delosSantos2011,FranzeseBiancoFoodBio2013,Bianco2014SR}.
We demonstrate how the cold- and $P$-denaturation mechanisms
 can emerge as a
competition between different free energy contributions coming from water, one
from  hydration water  and another from bulk water. Moreover, we show
how changes in the protein
sequence affect the hydration water properties and, in turn,
the stability of the protein folded state---a relevant information in
protein design \cite{Coluzza2011}.

The many-body  water model adopts a coarse-grain (CG) representation
of the water coordinates by partitioning  the
available volume $V$ into a fixed number $N_0$ of cells, each with
volume $v\equiv V/N_0\geq v_0$, where $v_0$ is the water 
excluded volume. 
Each cell accommodates at most one molecule with the average O--O distance between
next neighbor (n.n.) water molecules given by $r=v^{1/3}$.
To each cell we associate
a variable $n_{i}=1$ if the cell $i$  is occupied by a water molecule
and has $v_0/v> 0.5$, and $n_{i}=0$ otherwise.
Hence, $n_{i}$ is
a discretized density field replacing the water translational degrees
of freedom. 
The Hamiltonian of the bulk water 
\begin{equation}
\mathscr{H}\equiv \sum_{ij} U(r_{ij}) -J N_{\rm HB}^{\rm (b)}-J_\sigma 
N_{\rm  coop}
\label{bulk}
\end{equation} 
has a  first term, summed over all the water molecules $i$ and $j$ at
O--O distance $r_{ij}$, accounting 
for the van der Waals interaction, with $U(r)\equiv \infty$ 
for $r<r_0\equiv v_0^{1/3}= 2.9 $ \AA{} 
(water van der Waals
diameter), $U(r)\equiv 4\epsilon
[(r_0/r)^{12}-(r_0/r)^6]$ for $r\geq r_0$ with  $\epsilon \equiv 5.8$ kJ/mol
and $U(r)\equiv 0$ for $r>r_c \equiv 6 r_0$ (cutoff). 

The second term represents the directional component of the
hydrogen bond (HB), with 
$J/4\epsilon= 0.3$~\cite{note1},
$N_{\rm HB}^{\rm (b)}\equiv\sum_{\langle ij \rangle}n_i
n_j\delta_{\sigma_{ij},\sigma_{ji}}$ number of bulk HBs, with the sum over
n.n., where $\sigma_{ij}=1, \dots, q$ is the bonding index of molecule $i$ to the
n.n. molecule $j$, with $\delta_{ab}=1$ if $a=b$, 0 otherwise. Each water
molecule can form up to four HBs that break if $n_in_j=0$,
i.e. $r_{ij}>2^{1/3} r_0=3.6 $\AA{}, or ${\widehat{\rm OOH}}> 30^o$.
Hence only $1/6$ of 
the entire range of values $[0,360^\circ]$ for  the
${\widehat{\rm OOH}}$ angle is associated to a bonded state.
 Thus we choose $q=6$ to account correctly for the entropy
variation
due to  HB 
formation and breaking.

The third term,
 with $N_{\rm coop}\equiv \sum_i n_i\sum_{(l,k)_i}\delta_{\sigma_{ik},\sigma_{il}}$,    
 where $(l,k)_i$ indicates each of the six different pairs of the four
indices $\sigma_{ij}$ of a molecule $i$, 
accounts for  the HB cooperativity due to the quantum many-body interaction
\cite{Hernandez2005} and leads to the low-$P$ tetrahedral structure 
\cite{SoperRicci2000}.
We choose $J_\sigma/4\epsilon\equiv 0.05\ll J$, to guarantee an
asymmetry between the two HB terms.

Increasing $P$ partially disrupts the open structure of the HB network and reduces $v$
toward $v_0$. We account for this with an average enthalpy increase
$Pv_{\rm HB}^{\rm (b)}$ per HB,  where
$v_{\rm HB}^{\rm (b)}/v_0=0.5$ is the average volume increase between
high-$\rho$ ices VI and VIII and low-$\rho$ (tetrahedral) ice
Ih. Hence, the total bulk volume is 
 $V^{\rm (b)}\equiv Nv_0+N_{\rm HB}^{\rm (b)}v_{\rm HB}^{\rm (b)}.$ 
We assume that the HBs do not affect the n.n. distance $r$, consistent
with experiments \cite{SoperRicci2000}, hence do not affect the
$U(r)$ term. 

Next we  account for the effects of protein-water
  interaction.
Experiments and simulations 
 show that near a hydrophobic ($\Phi$) residue water-water HBs are
more stable then in bulk    \cite{Dias2008, petersonJCP2009, Tarasevich2011,  
  Davis2012}
with stronger  water-water correlation \cite{Sarupria2009}.   We model
this  by replacing $J$ of Eq.~(\ref{bulk}) with $J_{\Phi}>J$
for HBs at the $\Phi$ interface.
This choice, according to Muller \cite{muller1990},
ensures the water enthalpy compensation during the cold-denaturation
\cite{Bianco2012a}.

The  interaction at the $\Phi$ interface affects the 
hydration water density and fluctuations.  Some works
suggest a decrease of  interfacial water density \cite{LumKa1999,
  Schwendel2003, jensenPRL2003, DoshiPNAS2005}, while recent
simulations
show an increase of density in the first hydration shell of any
solute \cite{Godawat2009}  and an increase of 
compressibility near 
$\Phi$ solutes with size $\gtrsim$ 0.5~nm for water
\cite{Dadarlat2006, Sarupria2009,  Das2012} or water-like solvents
\cite{Strekalova2012} with respect to bulk.
Increasing $P$ induces a  further increase of density
and reduces the  effect of the $\Phi$ interface on the  compressibility of the
hydration shell \cite{Sarupria2009,Das2012,Ghosh2001,Dias2014}.   We
  incorporate this
  behavior  
into the model by using the following
considerations. From 
the equilibrium condition for the thermodynamic potential of
hydration water and the coexisting vapor at the $\Phi$ interface at
fixed $T$, according to the Eq. (2) of Ref. \cite{Giovambattista2006},  
we deduce
$v^{(\Phi)}-v_0 \sim (P-P^*)^{-1}$, where $v^{(\Phi)}$ is the 
volume per hydration water molecule and 
$P^*<0$ is the equilibrium
vapor pressure at the given $T$. If
we attribute this $P$-dependence to the interfacial HB properties
($v_{\rm HB}^{(\Phi)}\sim v^{(\Phi)}-v_0$) 
and expand it as a power series in $P$,
the average volume change
per water-water HB at the $\Phi$ interface is 
\begin{equation}\label{vsurf}
 v_{\rm HB}^{(\Phi)}/v_{\rm HB,0}^{(\Phi)}\equiv 1- k_1P + k_2 P^2
 -k_3 P^3+O(P^4)
\end{equation} 
where $v_{\rm HB,0}^{(\Phi)}$ is the volume change associated to the
HB formation in
the $\Phi$ hydration
shell at $P=0$, $k_i>0$ $\forall i$ and 
$\lim_{P\rightarrow \infty } v_{\rm HB}^{(\Phi)}=0$.
Hence,
the total volume $V$ is 
\begin{equation}
  V\equiv  V^{\rm (b)}
+  V^{(\Phi)} \equiv
  V^{\rm (b)}
  + N_{\rm HB}^{(\Phi)} v_{\rm HB}^{(\Phi)},
  \label{Vtot}
\end{equation}
where $V^{(\Phi)}$ and 
$N_{\rm HB}^{(\Phi)}$ are the $\Phi$ hydration shell volume and number of
HBs, respectively.

Because we are interested to small values of $P$, i.e. near the
biologically relevant atmospheric pressure, we include in our
calculations  only the linear term in Eq.(\ref{vsurf})~\cite{note2}.
In the following we fix 
$k_1= 1 v_0/4\epsilon$,
$v_{\rm HB,0}^{(\Phi)}/v_0=v_{\rm HB}^{\rm (b)}/v_0=0.5$ and
$J_{\Phi}/J=1.83$.
 Our results have minor qualitative differences
by
including up to the third order in Eq.(\ref{vsurf})  or
  by changing up to 50\% the parameters.

\begin{figure}
 \includegraphics[scale=0.4,bb=0 10 600 450,clip=true]{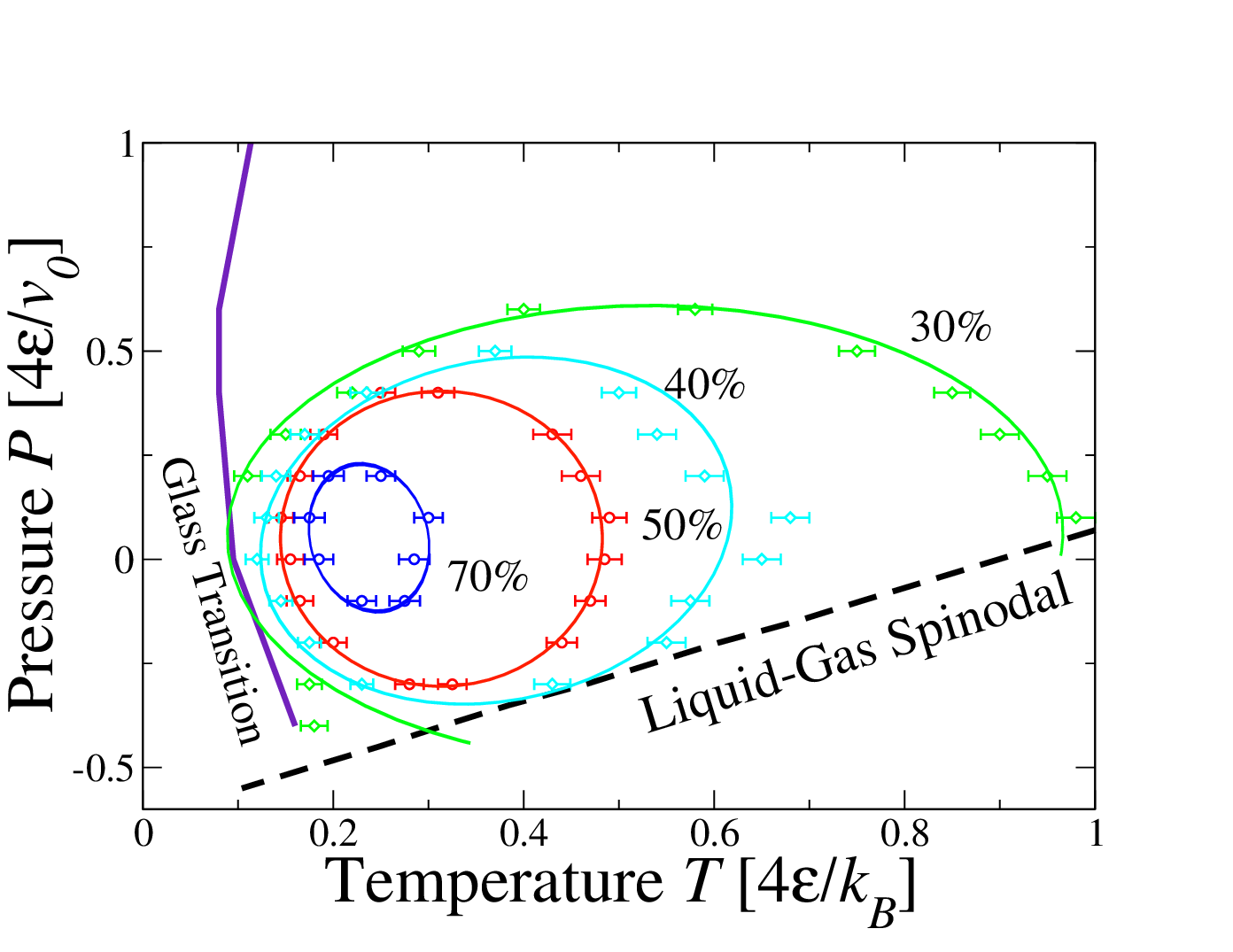}
\caption{ $P-T$ stability region (SR) of the protein from
  MC
  simulations.
   Symbols 
  mark state points with the same  
  average residue-residue contact's
  number $n_{\rm rr}/n_{\rm  max}=30\%$, 40\%, 50\% and 70\%.
Elliptic lines are guides for the eyes. The ``glass transition'' line
defines the temperatures below which the system does not 
  equilibrate. 
The spinodal line marks the stability limit of the liquid
phase at high $P$ with respect to the gas at low $P$; 
$k_B$ is
the Boltzmann constant.} 
\label{SR}
\end{figure}

Because our goal here is to calculate the water contribution to
denaturation, we model the protein as a self-avoiding $\Phi$
homopolymer,   without internal
  cavities \cite{note-cavity},
whose residues occupy n.n. cells with no residue-residue
interaction but the excluded volume, as in other  CG
approaches to the  problem
\cite{Marques2003,  Buldyrev2007, 
  Patel2007, Dias2008}.
This implies that the protein has several
``native'' states, all with the same maximum number $n_{\rm max}$ of residue-residue
contacts.  To simplify the discussion, we
  initially neglect energetic contributions of  
  the water-$\Phi$ residue interaction.

We analyze the system by Monte Carlo  (MC) simulations at constant $N$, $P$,
$T$. 
We adopt a representation in
two dimensions (2D) \cite{Marques2003, Patel2007, Dias2008,
  Lau2002, delosrios2000}, using a square partition, 
  to favor visualization and understanding 
  of our results.
Comparisons with our preliminary results in 3D
do not show 
qualitative changes, mainly because the number of n.n. 
water molecules is four both in 2D and 3D for the
tendency of water to form tetrahedral structures in 3D.

\begin{figure}
\centering
\includegraphics[scale=0.12,bb=0 -1 1848 4547, clip=true]{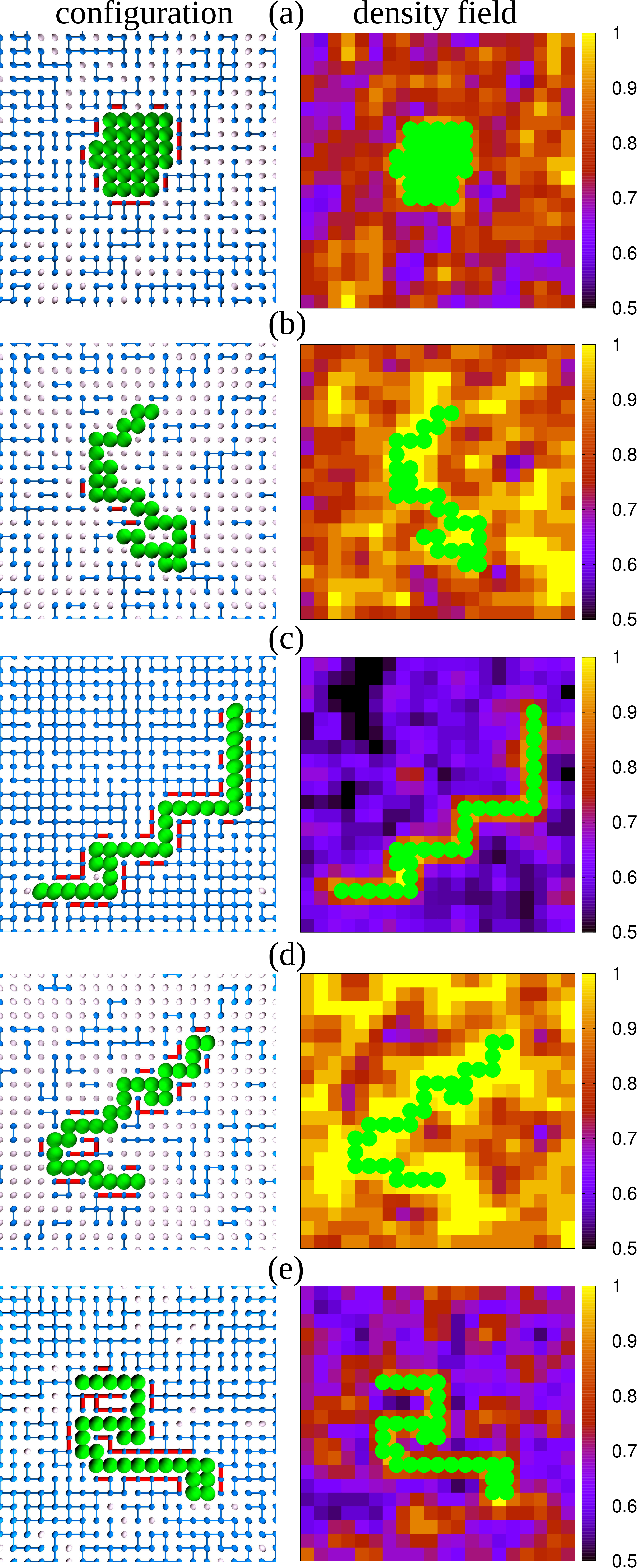}
\caption{Typical configurations 
  of a hydrated protein made of 30
  residues (in green): (a) folded at the state point 
  $(Tk_B/4\epsilon, Pv_0/4\epsilon)=(0.25, 0.1)$ and unfolded 
  (b) at high-$T$  $(0.9, 0.1)$;
  (c) at low-$T$  $(0.1, 0.1)$;
  (d) at high-$P$  $(0.25,  0.6)$;
  (e) at low-$P$ $(0.25, -0.3)$. 
  Left panels: Water molecules with/without HBs are represented
  in blue/white and bulk/interfacial HBs in 
  blue/red. Right panels:   Color coded water density field
  (from black for lower $\rho$ to yellow  for higher $\rho$) 
  calculated as
  $v_0\rho_i^{(\lambda)}\equiv v_0/(v_0+ n_{{\rm HB},i}^{(\lambda)} v_{\rm HB}^{(\lambda)})$ 
  where $\lambda = {\rm b}, \Phi$, and $n_{{\rm HB},i}^{(\lambda)}$ is
  the number of HBs associated to the water molecule $i$, with
  $\sum_i n_{{\rm HB},i}^{(\lambda)} = N_{\rm HB}^{(\lambda)}$.}
\label{prot_conf}
\end{figure}

We consider that the protein is folded if 
the average number of residue-residue contacts
$n_{\rm rr}\geq 50\% ~n_{\rm  max}$.
We find an elliptic SR (Fig.\ref{SR}), consistent with experiments and the Hawley
theory \cite{hawley1971,Smeller2002}, with 
heat-, cold-, and $P$-unfolding. The elliptic shape is preserved
when we change the threshold  for $n_{\rm rr}$,
showing that the $f \longleftrightarrow u$ is a continuous process.
In the SR the folded protein (Fig. \ref{prot_conf}a) 
minimizes the number of hydrated $\Phi$ residues, reducing
the energy cost of the interface, as expected.

First, we observe that the model reproduces the expected
{\em  entropy-driven} $f \longleftrightarrow u$ for increasing
$T$ at constant $P$
(Fig. \ref{prot_conf}b). The entropy $S$ increases both
for the opening of the protein and for the larger decrease of
 HBs.

Upon isobaric decrease of $T$ 
the internal energy dominates the system Gibbs free energy (Fig. \ref{prot_conf}c).
However, $N_{\rm  HB}^{\rm (b)}$ saturates at $T$ lower than the SR,
therefore the only way for the system to further minimize
the internal energy is to increase $N_{\rm  HB}^{(\Phi)}$, i.e. to
unfold the protein. Hence the cold denaturation is an {\em energy-driven}
process toward a protein state that is stabilized by the increased
number of HBs in the hydration shell.

Upon isothermal increase of $P$ the protein denaturates
(Fig. \ref{prot_conf}d).
 We find that this change is associated to a
decrease of $N_{\rm  HB}^{\rm (b)}$ and a small increase of
$N_{\rm  HB}^{\rm (\Phi)}$ that lead to a net decrease of $V$ at
high $P$, as a consequence of Eqs.~(\ref{vsurf}) and (\ref{Vtot}), and
an increase of internal energy. At high $P$
the $PV$-decrease associated to the $f \longrightarrow u$ process
at constant $T$ dominates over
the concomitant internal energy increase, determining a lower  
Gibbs free energy for the $u$ state. Hence water contribution to 
the high-$P$ denaturation is {\em density-driven}, as
emphasized by the increase of local density near the unfolded
protein.

Finally, upon 
isothermal decrease of $P$ toward negative values 
(Fig. \ref{prot_conf}e),
the enthalpy decreases when the contribution
$(Pv_{\rm  HB}^{(\Phi)}-J_\Phi)N_{\rm HB}^{(\Phi)}$ decreases,
i.e. when $N_{\rm HB}^{(\Phi)}$ increases. Therefore we find that
under depressurization the denaturation process 
is {\em enthalpy-driven}.

\begin{figure}
\includegraphics[scale=1,bb= 0 2 241 150,clip=true]{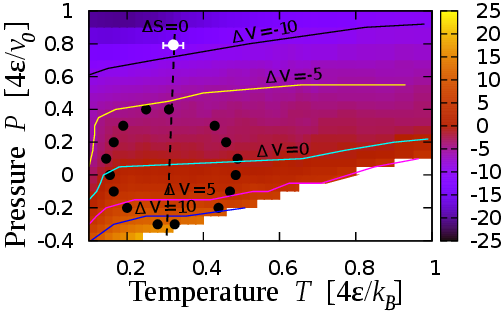}
\caption{Volume change $\Delta V$  for the
  $f \longrightarrow u$ process in the $T-P$ plane.
Color coded $\Delta V$
  (black for
  negative, yellow for positive) is in $v_0$ units. Solid lines
connect state points with constant   
  $\Delta V$.  Black points mark the SR. 
The locus $\Delta V=0$ has a positive slope and intersects the SR at
the turning points with  $dT/dP|_{\rm SR}=0$.
The dashed line, connecting the
points with $dP/dT|_{\rm SR}=0$,
corresponds to the locus where $\Delta S=0$
and separates state points with $\Delta S>0$ (high $T$) from those with  $\Delta S< 0$
(low $T$) at the $f \longrightarrow u$ process.
The white symbol marks the error in the dashed-line slope estimate.}
\label{deltaV_S}
\end{figure}

From the Clapeyron relation $dP/dT = \Delta S/ \Delta V$ applied to
the SR \cite{hawley1971}, we expect that
the $f \longleftrightarrow u$ process is isochoric
at the SR turning points where $\partial T /\partial P|_{\rm SR}=0$, 
while  is isoentropic
at the turning points where $\partial P/\partial T|_{\rm SR}=0$.
In particular, 
at any $T$ and $P$ 
the volume change in the $f \longrightarrow u$
process is given by
\begin{equation}\label{vol_surf}
  \Delta V \equiv V_u-V_f\simeq  v_{\rm HB}^{\rm (b)} \Delta N_{\rm HB}^{\rm (b)}  +
  (v_{\rm HB, 0}^{\rm(\Phi)} -k_1P)
  \Delta N_{\rm HB}^{\rm(\Phi)}. 
\end{equation}
We estimate the Eq. (\ref{vol_surf}) 
calculating the average volume $V_u$ and $V_f$ in a wide range of $T$ and $P$,
equilibrating water 
around a completely
unfolded protein state and a completely folded state (with
$n_{\rm rr}=n_{\rm max}$).
Consistently with the Hawley's theory  \cite{hawley1971}, we find that
the
$T$-denaturation   is accompanied by a positive entropy
variation $\Delta S>0$ at high $T$ 
and an {\em entropic penalty} $\Delta S<0$ at low $T$, while the
$P$-denaturation
  by a decrease of volume $\Delta V<0$  at
high $P$ and an increase of volume $\Delta V>0$ at low $P$
(Fig. \ref{deltaV_S}).
In particular, at
$P=0.3(4\epsilon/v_0)$, corresponding to $\approx 500$~MPa, we find
$\Delta V\approx -2.5 v_0$, hence
$|P\Delta V|=0.75 (4\epsilon)\approx 17$~kJ/mol, very close to the
typical reported value of 15~kJ/mol \cite{MeersmanFilip2006}.
By varying the parameters $v_{\rm HB}^{(\Phi)}$ and $J_{\Phi}$
we find that the first is relevant for the $P$-denaturation, as expected
because it dominates Eq.~(\ref{Vtot}), while the second affects the
stability range in $T$. Both combine in a non-trivial way to regulate
the low-$T$ entropic penalty. 
  We test our results including a small water-$\Phi$ residue attraction and find no 
qualitative differences but  a  small change in the $T$-range of
stability of  the folded protein.

\begin{figure}
\includegraphics[scale=0.4,bb=0 10 668 480,clip=true]{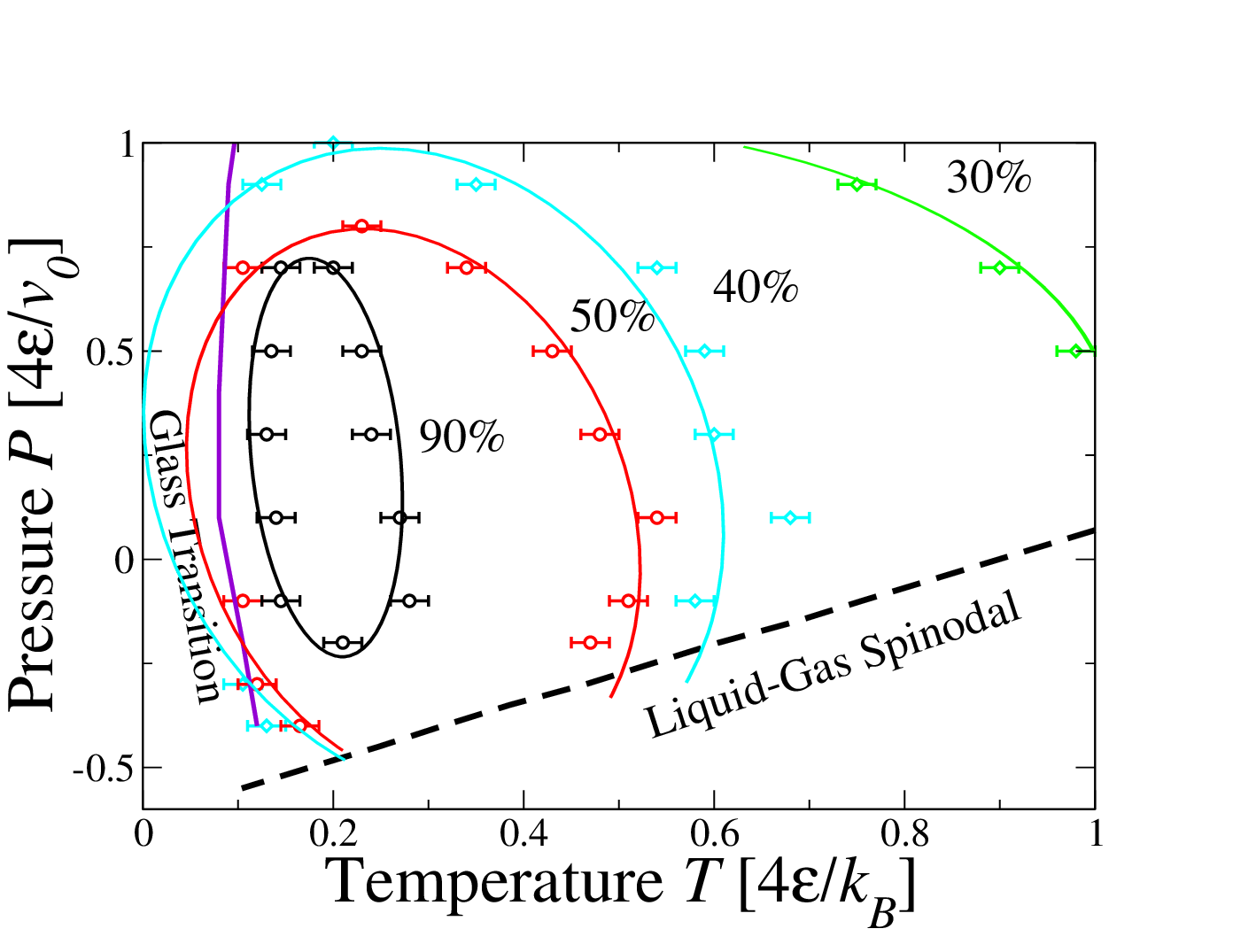}
\caption{The SR for the  heteropolymer with a unique native state.
  We set
  $\epsilon_{\rm rr}/J=0.7$,
  $\epsilon_{\rm w,\Phi}=0$,  
  $\epsilon_{\rm w,\zeta}/J=1.17$,
  $J_{\Phi}/J=1.3$,
  $J_{\zeta}/J=0.5$,
  $v_{\rm  HB}^{(\zeta)}=0$,
  with all the other parameters as in
  Fig.~\ref{SR}. We test that changing the parameters,
  within physical ranges,
  modifies the SR, reproducing a variety of experimental
 SRs \cite{Smeller2002},  but  preserves the elliptic shape.} 
\label{SR1}
\end{figure}

Next, 
we study the case
of a protein with hydrophobic ($\Phi$) and
hydrophilic ($\zeta$) residues \cite{Lau2002,delosrios2000},
  with a residue-residue interaction
matrix $A_{i,j}=\epsilon_{\rm rr}$ if residues $i$ and $j$ are
n.n. in the unique native state, $0$ otherwise. Water molecules interact with
energy $\epsilon_{\rm w,\Phi}<J$ 
and $\epsilon_{\rm w,\zeta}>J$ with n.n.  $\Phi$ and  $\zeta$
residues respectively, accounting for the polarization of the
solvent near the $\zeta$
residues. The polar $\zeta$ residues
 disrupt the tetrahedral  order of the
surrounding water molecules.  Thus we assume that a $\zeta$ residue
$j$ and a n.n. water molecules $i$ form a HB
when the latter has $\sigma_{i,j}$ in the state 
$q_j^{(\zeta)}=1, \dots, q$ preassigned to $j$.
Finally, we consider that water-water enthalpy in
the hydration shell is
$H_{\lambda, \lambda}\equiv -J_{\lambda} + Pv^{(\lambda)}_{\rm  HB}$,
if both molecules are n.n. to the same type of residue
or $H_{\lambda, \mu}\equiv (H_{\lambda, \lambda} + H_{\mu, \mu})/2$ if
the n.n. residues belong to different types,
with $\lambda, \mu = \Phi, \zeta$, and 
$J_{\zeta}\leq J$ \cite{Cheng2003}
(Fig.\ref{SR1}).

 Despite the complexity of the  heteropolymer model,
  we find results that are similar to the homopolymer case, with the
  qualitative difference  that with $\zeta$-residues we find 
    a locus $\Delta
  V=0$ with negative slope and increased stability 
  toward (i) cold- and (ii)  $P$-denaturation.
    In particular, for our
  specific choice of parameters, for the heteropolymer the cold denaturation at $P=0$ occurs
  below the glass transition, instead of 
  $\approx 50\%$ above as for the homopolymer.
  Furthermore,  the SR against $P$ is $\approx 2$ times larger with $\zeta$-residues
  than without.
This comparison suggests that the water
contribution is relevant to the $f \longleftrightarrow u$
independently on the residue sequence, although the
residue-residue interactions increase the stability of the folded state.
%

In conclusion, our model for protein folding 
reproduces the entire protein SR in explicit solvent and allows us to
identify how water contributes to the $T$- and $P$-denaturation
processes. The model is thermodynamically consistent with 
Hawley’s theory but, in addition, allows for intermediate states for
the $f \longleftrightarrow u$  process. We find 
that cold denaturation is energy-driven, while unfolding by
pressurization and depressurization, in addition to other
  suggested mechanisms \cite{RochePNAS2012},  are density- and enthalpy-driven
 by water, 
respectively. For these mechanisms is essential to take into
account how  the
  protein-water interactions affect the stability of 
 the water-water HB and the water density
in the hydration shell. In particular, both
properties control the low-$T$ entropic penalty.
Our results are qualitatively robust against
modification of the model parameters, within physical ranges, and the
model is computationally efficient thanks to the adoption of a CG
water model,  
representing a step towards the development of a theoretical and
computational approach for protein design and engineering.

\bigskip

We thank
M. Bernabei,
I. Coluzza,
C. Karner,
E. Locatelli,
P. Malgaretti
and
N. Patges
for discussions, 
Spanish
FIS2012-31025 and EU
NMP4-SL-2011-266737 grants for
support. V.B. acknowledges support from Catalan 
FI-DGR 2010 and Italian
``Angelo della Riccia''
grants. 

\bigskip

\bibliographystyle{apsrev4-1}

\begin{thebibliography}{99}
\bibitem{note2}
  This
approximation implies that our calculations are valid only for
$P<1/k_1$ where the HB contribution to the isothermal compressibility of the $\Phi$
hydration shell, $ K_T^{({\rm HB},\Phi)}\equiv - (1/V^{(\Phi)}) (\partial
V^{(\Phi)}/\partial P)_T\simeq  (k_1/(1-k_1 P)) - (1/N_{\rm HB}^{(\Phi)}) (\partial N_{\rm HB}^{(\Phi)}/ \partial P)_T$
is finite.
\end{thebibliography}



\end{document}